\documentclass[a4paper,12pt]{article}

\newcommand{\sect}[1]{\setcounter{equation}{0}\section{#1}}

\newcommand{\bea}{\begin{eqnarray}}
\newcommand{\eea}{\end{eqnarray}}
\newcommand{\be}{\begin{equation}}
\newcommand{\ee}{\end{equation}}
\newcommand{\vs}[1]{\vspace{#1 mm}}

\newcommand{\dsl}{\pa \kern-0.5em /}

\newcommand{\pa}{\partial}

\begin{document}
\topmargin 0pt
\oddsidemargin 0mm

%\renewcommand{\thefootnote}{\fnsymbol{footnote}}
%\begin{titlepage}
\begin{flushright}
hep-th/0202088\\
%SINP-TNP/02-? 
\end{flushright}

\vs{2}
\begin{center}
{\Large \bf  
A Short Note On The Wilson-Loop Average And The AdS/CFT Correspondence.}

\vs{10}

{\large Somdatta Bhattacharya}
\vspace{5mm}

{\em 
 Saha Institute of Nuclear Physics,
 1/AF Bidhannagar, Calcutta-700 064, India\\
E-Mails: som@theory.saha.ernet.in\\}
\end{center}

\vs{5}
\centerline{{\bf{Abstract}}}
\vs{5}
\begin{small}
In hep-th/9803002, Maldacena argued that in the light of the AdS/CFT
correspondence as formulated by Witten and Gubser, Klebanov and Polyakov
as a relation between partition functions, the expectation value of
the Wilson Loop in \( \mathcal{N} \)=4 SU(N) SYM is given by the
world sheet partition function with the action formulated on an 
\( AdS_{5}\times S^{5} \)
background and the world sheet ending on the loop on the boundary
of \( AdS_{5}. \) What we propose to do in this paper is to give
some alternative arguments as to why it should be so.\newpage
\end{small}
%% LyX 1.1 created this file.  For more info, see http://www.lyx.org/.
%% Do not edit unless you really know what you are doing.
%\documentclass[12pt,english,english,english]{article}
%\usepackage[T1]{fontenc}
%\usepackage[latin1]{inputenc}
%\usepackage{babel}

%\makeatletter

%%%%%%%%%%%%%%%%%%%%%%%%%%%%%% LyX specific LaTeX commands.
%\providecommand{\LyX}{L\kern-.1667em\lower.25em\hbox{Y}\kern-.125emX\@}

%%%%%%%%%%%%%%%%%%%%%%%%%%%%%% User specified LaTeX commands.
%\usepackage[T1]{fontenc}
%\usepackage[latin1]{inputenc}
%\usepackage{babel}

%\makeatletter

%\usepackage[T1]{fontenc}
%\usepackage[latin1]{inputenc}
%\usepackage{babel}

%\makeatletter

%\makeatother

%\makeatother

\sect{Introduction}

The AdS/CFT correspondence \cite{mald1,gkp,witt} has come a long way in
 elucidating
the long conjectured duality between gauge theories and string theories.
The case most studied and with respect to which the most evidence
has accumulated is the case of \( \mathcal{N} \)=4 SU(N) Super Yang
Mills being dual to type IIB string theory on \( AdS_{5}\times S^{5} \)
[\cite{agmoo} and references therein]. However, there is a caveat on most
of these evidences, because the duality relates the weakly coupled
regime of one theory to the strongly coupled one of the other and
vice versa. For instance, matching of the correlation functions on
both sides makes use of non-renormalization theorems, which keeps
one wondering whether these are really evidences or are dictated by
symmetry alone.

Gauge theory, however, has a formal non-perturbative formulation,
in terms of the loop equation \cite{mm}, which allows one to define if
not compute quantities at arbitrarily large couplings. The loop equation
is an equation for the expectation value of the Wilson loop and it
is argued that the following ansatz is the solution \cite{mald2,dgt}

\begin{equation}
\left\langle W\right\rangle =\int \frac{DXDgD\theta }{Vol(diff\times Weyl)}
e^{-S}
\end{equation}
 where \( S \) is the Green-Schwarz superstring action on an 
\( AdS_{5}\times S^{5} \)
background as has been formulated in \cite{mt} and the path integral
is over all embeddings of the string into \( AdS_{5}\times S^{5} \)
with proper boundary conditions (the string world surface should end
along the loop at the boundary of \( AdS_{5} \)).

Here \( W \) is given by

\begin{equation}
\label{eq0}
W=\frac{1}{N}TrPexp(i\oint (A_{\mu }\dot{X}^{\mu }-i|\dot{X}|\Phi _{i}
\theta ^{i})ds)
\end{equation}
 where \( A_{\mu } \) are the gauge fields and \( \Phi _{i} \) are
the six scalars in the adjoint representation. The \( i \) in front
of the second term in the integral is due to the metric being Euclidean.
The \( \theta  \) here are not the same as the \( \theta  \) in
the Green Schwarz action. They are angular coordinates of magnitude
1 and can be regarded as coordinates on \( S_{5} \). The former denote
the Green Schwarz fermions.

Polyakov has argued for such a relationship to hold for the case of
the non-supersymmetric Yang Mills with the corresponding string path-integral
being that for just \( AdS_{5} \), from a totally different viewpoint
\cite{poly}. Moreover, Maldacena \cite{mald2} has given some arguments based
on the AdS/CFT correspondence in the form as given in \cite{gkp,witt}. See
\cite{rey} for related developments.

What we propose to do is to forward some alternative arguments as
to why such a relationship should follow from the AdS/CFT correspondence.
Our arguments will be based upon certain reasonable assumptions which
are hard to prove and we will not try to prove them here but simply
assume them to be true. We will point out where we will make these
assumptions.

\sect{The Arguments}

On one side of the AdS/CFT correspondence, one has the string theory
partition function, and on the other side one has the Yang-Mills partition
function. We shall be interested in the Yang-Mills partition function
without any sources. On the string theory side, this means that no
fields are kept fixed on the boundary of \( AdS_{5} \) which is \( R^{4} \).

Thus the string theory partition function is

\begin{equation}
Z=\int D\phi _{i}e^{-S[\phi _{i}]}
\end{equation}

where \( \phi _{i} \) are the various fields in the entire spectrum
of type IIB superstring theory and the action is formulated about
the \( AdS_{5}\times S^{5} \) background.

Now it is well known that the one-loop free energy for conformal backgrounds 
(in the sense of world-sheet conformal invariance)
is just the string world sheet partition function on the torus, formulated
with the background acting as the sigma model couplings, i.e., for
bosonic string theories,

\begin{equation}
F=lnZ=\int _{torus}\frac{DXDg}{Vol(diff\times Weyl)}e^{-[G_{MN}g^{ab}
\partial _{a}X^{M}\partial _{b}X^{N}+....]}
\end{equation}
 where \( G_{MN} \) is part of the conformal background.

Thus, the free energy for type IIB superstring theory on the \( AdS_{5}\times
 S^{5} \)
background (which is a conformal background as the \( AdS_{5}\times S^{5} \)
background is a solution of the supergravity equations of motion)
can be written as \begin{equation}
\label{eq1}
F=\int _{torus}\frac{DXDgD\theta }{Vol(diff\times Weyl)}e^{-S_{G.S.}[X^{M},
g^{ab},\theta ]}
\end{equation}

where \( S_{G.S.}[X^{M},g^{ab},\theta ] \) is the Green-Schwarz action
for type IIB superstrings on an \( AdS_{5}\times S^{5} \) background
as has been formulated by Metsaev and Tseytlin \cite{mt}. The RR fiveform
is incorporated into one of the fermionic terms in the action. This
is assumption no.1 as there is no well-defined quantum theory as yet
of the Green-Schwarz action formulated over the \( AdS_{5}\times S^{5} \)
background. However if the Metsaev-Tseytlin action is the correct
world sheet action for the Type IIB theory on an \( AdS_{5}\times S^{5} \)
background then such a relationship between the one-loop free energy
and the torus partition function has to hold.

Now the Yang-Mills partition function can be reformulated in terms
of the master variables which can be taken as the Wilson loops. What
the exact path integral formulation will be is not known but we will
not need the exact form here. We will assume that the the YM partition
function is expressible in the form\begin{equation}
\label{eq2}
Z=\int D\Psi e^{-S[\Psi ]}
\end{equation}

where \( \Psi =TrPe^{i\oint (A_{\mu }\dot{X}^{\mu }-i|\dot{X}|\Phi _{i}
\theta ^{i})ds} \)
and \( S[\Psi ] \) is some unknown functional in terms of the \( \Psi  \)
variables, which includes the Jacobian factor in transforming from
the \( A_{\mu } \) to the \( \Psi  \) variable.

The point to be noted is that since the \( \Psi  \) variable is a
function of the curve \( (X^{\mu }(s),\theta ^{i}(s)) \), it is like
a string field in 10-d. That a Wilson loop can be interpreted as a
string field had been conjectured in the early eighties \cite{mm} and 
[\cite{neveu} and references therein]. In the more modern treatment of
 Polyakov \cite{poly},
such a relationship is implicit. Also, such a relationship has been
used in \cite{gid}. If we denote \( |\dot{X}|\theta _{i} \) by \( Y_{i} \)
, the resultant 10-d coordinates obey \( \dot{X}^{2}=\dot{Y}^{2}. \)
Thus the metric in this space has signature (4,6) and is hence not
the signature of \( AdS_{5}\times S^{5}, \) but of the space where
the loops are defined. As has been shown in \cite{dgo}, this is related
to the fact that the six loop variables \( Y_{i}(s) \) correspond
to T-dual coordinates on the string worldsheet. Loops obeying this
condition are BPS loops as they are invariant under half the supersymmetries
in super-loop space. We refer the reader to \cite{dgo} for further details.
The resulting interpretation is thus that if the AdS/CFT correspondence
is correct, the string field theory action corresponding to type IIB
string theory on an \( AdS_{5}\times S^{5} \) background is given
by the unknown action functional \( S[\Psi ] \) in (\ref{eq2}).
For all practical purposes, it is an interacting string field theory
incorporating splitting-joining interactions and zig-zag symmetry.
And according to this interpretration, these interactions are precisely
taken care of by the non-trivial background on the string side. This
is assumption no.2 and it is a very strong one but it would hold if
indeed the \( \Psi  \) can be assumed to be a string field and if
the AdS/CFT correspondence is valid.

Thus if this interpretation is correct, the free energy corresponding
to this string field theory partition function as obtained through
the AdS/CFT correspondence is given by (\ref{eq1}).

Now it is a well known fact in string field theory that given a torus
partition function for the free energy, the correlator for two string
fields/Wilson Loops is given by the partition function over a cylinder
with the same action, with the strings forming the boundary of the
cylinder acting as the coordinates for the string fields, i.e.

\[
\left\langle \Psi (C_{f})\Psi (C_{i})\right\rangle =\int _{C_{i}}^{C_{f}}
\frac{DXDgD\theta }{Vol(Diff\times Weyl)}e^{-S[X^{M}(\sigma ,\tau ),g^{ab}
(\sigma ,\tau ),\theta _{\alpha }(\sigma ,\tau )]}\]

In the above \( C_{f} \) and \( C_{i} \) are both parametrised by
some \( X_{\mu }(s) \) and \( \theta _{i}(s) \).

Once the propagator is known, the vev for a single field follows in
a straightforward fashion. It is just the {}``one-point'' function
and is hence given by the same partition function over a disc, with
the boundary being formed by the coordinates of the string field/Wilson
loop whose vev is being taken, i.e.\[
\left\langle \Psi (C)\right\rangle =\int _{C}\frac{DXDgD\theta }
{Vol(diff\times Weyl)}e^{-S[X^{M},g^{ab},\theta _{\alpha }]}\]

This is the expression for the Wilson Loop average that appears in
\cite{dgt}. When evaluated it yields the area of the minimal surface
bounded by the loop in the classical limit, calculated with the
 \( AdS_{5}\times S^{5} \)
metric.

\sect{Discussion}

Thus in this note we have shown how under reasonable assumptions based
on the AdS/CFT correspondence the Wilson loop average can be shown
to be the world sheet partition function on an \( AdS_{5}\times S^{5} \)
background bounded by the loop which in the classical limit yields
the exponential of the minimal surface area bounded by the loop. In
the process we have conjectured an equivalence between the Yang-Mills
partition function and the type IIB string field theory partition
function on an \( AdS_{5}\times S^{5} \) background. Though such
a relationship is hard to prove, given the non-availability of a path-integral
formulation of Yang-Mills theory in terms of Wilson loops, it appears
that once available it would have to conform to such a relationship
in the face of the AdS/CFT correspondence. Polyakov has conjectured
that the relationship between the Wilson loop average and the world
sheet partition function bounded by the loop is true for the non-supersymmetric
case, from a totally different perspective. He has argued that the
solution of the loop equation which is an equation of the loop average
should yield the world sheet partition function on an \( AdS_{5} \)
background bounded by the loop and hence the exponential of the
area of the minimal surface bounded by the loop \cite{poly}. He and Rychkov
have shown that the above is true in the {}``WKB approximation''
\cite{pr}. Drukker, Gross and Ooguri have worked with the \( N=4 \)
case and have shown that for the above mentioned BPS loops, the above
holds good to a certain extent, in the sense that the loop average
is given by the minimal surface for smooth non-intersecting loops \cite{dgo}.
 If indeed the Wilson loop average can be shown to be given
by the minimal area surface for all kinds of loops, we can turn the
previous arguments around to reason that if the expectation value
of the Wilson loop is indeed given by the minimal area surface in
\( AdS_{5}\times S^{5} \), then the relationship between the partition
functions also holds good via the conjectured relationship between
string theory partition function on \( AdS_{5}\times S^{5} \) and
the string field theory interpretation coming from the Yang-Mills
side. This would then constitute some evidence for the AdS/CFT correspondence
beyond the supergravity approximation.

\section*{Acknowledgements}

We wish to thank Satchidananda Naik and Shibaji Roy for reading the
manuscript and offering helpful suggestions.

\end{document}